\documentclass[twoside]{ilcws07}
\usepackage[latin1]{inputenc}
\usepackage[dvips]{graphicx,epsfig,color}
\usepackage{wrapfig,rotating}
\usepackage{amssymb,amsmath,array}

\begin{document}

\title{
Effective Theory Approach to\\ $W$-Pair Production near Threshold 
\thanks{
Talk given at  the 
International Linear Collider Workshop (LCWS/ILC07),
30 May - 3 Jun 2007, Hamburg, Germany.
}
} 
\author{Christian Schwinn
\vspace{.3cm}\\
RWTH Aachen - Institut f\"ur Theoretische Physik E\\  
D--52056 Aachen - Germany
}

\maketitle


\begin{abstract}
  In this talk, I review the effective theory approach to unstable
particle production and present results of a calculation of the
process $e^{-} e^{+} \rightarrow \mu^{-} \bar{\nu}_{\mu} u \bar{d}\,X$
near the $W$-pair production threshold up to next-to-leading order in
$\Gamma_W/M_W\sim \alpha\sim v^2$. The remaining theoretical
uncertainty and the impact on the measurement of the $W$ mass is
discussed.
\end{abstract}

\begin{center}
\texttt{SFB/CPP-07-44, 	arXiv:0708.0730 [hep-ph],
 August 6, 2007}
\end{center}

\section{Introduction}
  
The masses 
of particles  like the top quark, the $W$ boson 
or yet undiscovered particles like supersymmetric partners
can be measured precisely using threshold scans at an
$e^-e^+$ collider.
In particular the error of the $W$ mass could be reduced to 
 $6\,$MeV  by measuring
 the four fermion production cross section near the $W$-pair
threshold~\cite{Wilson:2001aw},
provided theoretical uncertainties are reduced well below $1\%$.
In such precise calculations one has to treat
 finite width effects systematically and without violating gauge invariance.
The next-to-leading order~(NLO) calculations of $W$-pair
production~\cite{Denner:1999kn}
available at LEP2 were done in the pole
scheme~\cite{Stuart:1991xk} and were supposed
to break down near threshold. 
The recent computation of the complete NLO corrections to $e^-e^+\to 4 f$
processes in the complex mass scheme~\cite{Denner:2005es}
is valid 
near threshold and in the continuum,  but is technically
demanding and required to compute
one loop six-point functions. 

In this talk, I report on the NLO corrections to the total cross section of the process
\begin{equation}
\label{eq:wwprocess}
e^-  e^+ \to \mu^-  \bar{\nu}_\mu  u  \bar{d} X
\end{equation}
near the $W$-pair
threshold~\cite{Beneke:2007zg} obtained using
effective field theory~(EFT)
 methods~\cite{Chapovsky:2001zt,Beneke:2003xh,Beneke:2004xd}.
This calculation is simpler than
the one of~\cite{Denner:2005es}
and results in an almost analytical expression of the result
that allows for a detailed  investigation of theoretical uncertainties.
However, the method is not easily extended to differential 
cross sections.
Section~\ref{sec:eft} contains the leading order~(LO) EFT
description while the NLO approximation
of the tree and the radiative corrections are
 described in Sections~\ref{sec:born} and~\ref{sec:nlo}, respectively. 
Results are presented in Section~\ref{sec:results} together with 
 an estimate of the remaining theoretical uncertainties  and a 
 comparison  to~\cite{Denner:2005es}.

\section{Unstable particle effective theory}
\label{sec:eft}
To provide a systematic treatment of finite width effects,
in~\cite{Chapovsky:2001zt,Beneke:2003xh} EFT methods were used to
expand the cross section simultaneously in the coupling constant
$\alpha$, the ratio $\Gamma/M$ and the virtuality of the resonant
particle $(k^2-M^2)/M^2$, denoted collectively by $\delta$.  The modes
at the small scale $\delta$ (the resonance, soft or Coulomb photons,
\dots) and the external particles are described by an effective
Lagrangian $\mathcal{L}_{\text{eff}}$ that contains elements of heavy
quark effective theory or non-relativistic QED and soft-collinear
effective theory~(SCET) (for reviews of the various EFTs see e.g
\cite{Rothstein:2003mp}).  ``Hard'' fluctuations with virtualities $
\sim M^2$ are not part of the EFT and are integrated out. Their effect
is included in short-distance coefficients in
$\mathcal{L}_{\text{eff}}$ that can be computed in fixed-order
perturbation theory without resummations of self-energies.  Finite
width effects are relevant for the modes at the small scale and are
incorporated through complex short-distance coefficients in
$\mathcal{L}_{\text{eff}}$ ~\cite{Beneke:2003xh,Hoang:2004tg}.

It might be useful to compare the EFT approach to the pole scheme for
the example of the  production of a single resonance $\Phi$ in the
inclusive process $f_1\bar f_2\to  X$.  The pole scheme provides a
decomposition of the amplitude into resonant and non-resonant
pieces~\cite{Stuart:1991xk}:
\begin{equation}
\label{eq:pole-scheme}
\mathcal{A}(s)|_{s\sim M^2}=
\frac{\mathcal{R}(\bar s)}{s-\bar s}+\mathcal{N}(s),
\end{equation}
where both $\bar s$, the complex pole of the
propagator defined by $\bar s-M^2-\Pi(\bar s)=0$, and $\mathcal{R}(\bar s)$, the
residue of $\mathcal{A}(s)$ at $\bar s$, are gauge independent.
In the EFT, it is convenient to 
obtain the cross section from the imaginary part of the
forward-scattering amplitude
that reads~\cite{Beneke:2003xh}
\begin{equation}
\label{eq:eft-amp}
i \mathcal{A}(s)|_{s\sim M^2}=  \int d^4 x \,
\langle f_1\bar f_2 |
T\left[i {\cal O}^\dagger_{\Phi f_1\bar f_2}(0)
i{\cal O}_{f_1\bar f_1\Phi}(x)\right]|f_1\bar f_2\rangle
+ \langle f_1\bar f_2 |i {\cal O}_{4f}(0)|f_1\bar f_2 \rangle\, .
\end{equation}
Here $ {\cal O}_{f_1\bar f_2\Phi}$  describes the production of  
$\Phi$  while 
 $ {\cal O}_{4f}$ describes  non-resonant contributions.
The matching coefficients of these operators 
are gauge independent since they are
computed from on-shell scattering amplitudes in
the underlying theory, where for unstable particles 
``on-shell'' implies $k^2=\bar s$.
The  structure of~\eqref{eq:eft-amp} is similar to~\eqref{eq:pole-scheme},
 but the
 EFT provides a field
theoretic definition of the several terms. 
 Higher order
corrections to the matching coefficients correspond to the
\emph{factorizable corrections} in the pole scheme. Loop
corrections to the matrix elements in the EFT 
correspond to the \emph{non-factorizable
corrections}~\cite{Chapovsky:2001zt}.

Turning to $W$-pair production near threshold, the appropriate 
effective Lagrangian to describe
the two non-relativistic $W$ bosons
with
$  k^2-M_W^2 \sim M_W^2 v^2
\sim M_W^2\delta $
is given by~\cite{Beneke:2004xd}
\begin{equation}
{\cal L}_{\rm NRQED} = \sum_{a=\mp} \left[\Omega_a^{\dagger i} \left(
i D^0 + \frac{\vec{D}^2}{2 {M}_W} - \frac{\Delta}{2} \right)
\Omega_a^i
+  \Omega_a^{\dagger i}\,
\frac{(\vec{D}^2-M_W \Delta)^2}{8 M_W^3}\,
\Omega_a^i\right]
\label{LNR}
\end{equation}
with the matching coefficient~\cite{Beneke:2003xh} $
\Delta \equiv (\bar s-M_W^2)/M_W$.  If 
$M_W$ is the pole mass, this
becomes $\Delta=-i \Gamma_W$.  
The fields 
 $\Omega^i_\pm\equiv \sqrt {2M_W} W^i_\pm$ 
describe the three physical polarizations of the $W$s; 
the unphysical modes are not part of the EFT~\cite{Beneke:2004xd}.
The covariant derivative $D_\mu \Omega_\pm^i \equiv
(\partial_\mu\mp i e A_\mu) \Omega_\pm^i$ 
includes  interactions with those photon 
 fluctuations  that 
keep the virtualities of the $\Omega$s
at the order $\delta$.
 These are \emph{soft} photons 
with $(q^0,\vec q)\sim (\delta,\delta)$  and 
 \emph{potential} (Coulomb) photons with  $(q^0,\vec q)\sim (\delta,\sqrt \delta)$.
Collinear photons are also part of the EFT but do not contribute at NLO.
The Lagrangian~\eqref{LNR} reproduces the expansion of the 
 resummed transverse $W$ propagator
in $\delta$, as can be seen by  writing the $W$ four-momenta  as
$k^\mu =M_W\, v^\mu + r^\mu$  with  $ v^\mu\equiv(1,\vec{0})$
and a potential residual momentum 
$(r^0,|\vec{r}|) \sim M_W\,(v^2, v)\sim  (\delta,\sqrt \delta)$:
\begin{equation}
\label{eq:wprop}
  \frac{i}{k^2-  M_W^2-\Pi^{W}_T(k^2)}
\left(-g_{\mu\nu}+\frac{k_\mu k_\nu}{k^2}\right)
  \qquad \Rightarrow\qquad
  \frac{i (-g_{\mu\nu}+v_\mu v_\nu)}{
2M_W (r^0-  \frac{{\vec r}^2}{2 M_W}+\frac{\Delta}{2}    )}\, .
\end{equation}
Higher orders
in the expansion of the propagator are reproduced by the higher order
kinetic terms in~\eqref{LNR} and
residue factors  included  in the 
production operators~\cite{Beneke:2007zg}.

The production of a pair of non-relativistic $W$ bosons is
described by the operator~\cite{Beneke:2004xd}
\begin{equation}
\label{LPlead}
{\cal O}_p^{(0)} = \frac{\pi\alpha}{s_w^2 M_W^2}
\left(\bar{e}_{c_2,L} \left(\gamma^i n^j+\gamma^j n^i\right)  
e_{c_1,L} \right)
\left(\Omega_-^{\dagger i} \Omega_+^{\dagger j}\right) 
\end{equation}
that is determined
from the on-shell tree-level scattering amplitude $e^-e^+\to W^+W^-$:
\begin{equation}
\label{eq:tree-match}
\parbox{90mm}{
   \includegraphics[height=14.5mm]{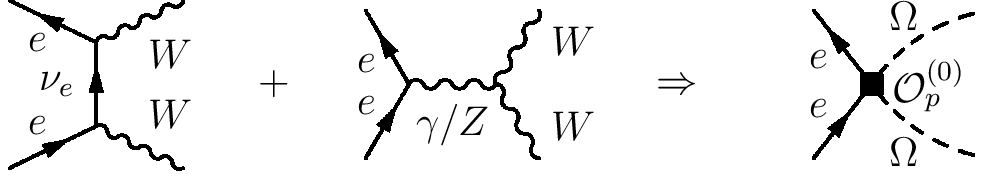}
}\, .
\end{equation}
At threshold, only the $t$-channel diagram and the $e^-_L e^+_R$ helicity
 contribute at leading order in $\delta$.
Similar to~\eqref{eq:eft-amp}, 
the LO $e^-e^+$ forward-scattering amplitude 
  in the EFT is given by the expectation value
of a time ordered product of the operators~\eqref{LPlead},
 evaluated using~\eqref{LNR}:
\begin{equation}
i {\cal A}^{(0)}=\int d^4 x\,
\langle e^- e^+|{\mathrm T}\lbrack i {\cal O}_p^{(0)\dagger }(0) i 
{\cal O}_p^{(0)}(x)\rbrack|e^- e^+\rangle=
\parbox{26mm}{
\includegraphics[height=15.5mm]{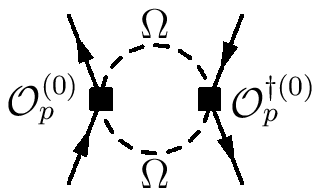}
 }\, .
\end{equation}
One estimates ${\cal A}^{(0)}\sim \alpha^2\sqrt{\delta}$,
 noting that each $\Omega$ 
propagator~\eqref{eq:wprop} contributes $\delta^{-1}$ and counting the
potential loop integral  as 
$dk^0 d^3 k_i\sim  \delta^{5/2}$.
The total cross section for the process~\eqref{eq:wwprocess} 
is obtained from
appropriate cuts  of  ${\cal A}^{(0)}$, 
where cutting  an  $\Omega_\pm$ line
 has to be interpreted as  cutting 
the self-energies
 resummed in the EFT propagator.
At LO, the cuts contributing to the flavour-specific 
final state are correctly extracted by 
multiplying the imaginary part of ${\cal A}^{(0)}$ 
by the leading-order branching
fractions. In terms of $E=\sqrt s-2M_W$ one obtains~\cite{Beneke:2007zg} 
\begin{equation}
\label{eq:sigmaLO}
\sigma^{(0)}(e^-  e^+ \to \mu^-  \bar{\nu}_\mu  u  \bar{d})
= \frac{\pi \alpha^2}{27 s_w^4 s}\text{Im}\left[- i \,\sqrt{-\frac{E+i \Gamma_W}{M_W}}\;\right] \, .
\end{equation}

\section{NLO EFT approximation to the born cross section}
\label{sec:born}

Some  parts of the NLO EFT calculation 
of the process~\eqref{eq:wwprocess}
are included in a Born calculation in
the full theory with a fixed width prescription.
One contribution arises from \emph{four-electron operators}
 ${\cal O}^{(k)}_{4e}$
 analogous to those in~\eqref{eq:eft-amp}.
Their matching coefficients $C^{(k)}_{4e}$
are obtained from the forward-scattering
amplitude  in the full electroweak theory.
The leading 
imaginary parts of $C^{(k)}_{4e}$ are of order $\alpha^3$ and
arise from cut two-loop diagrams
corresponding to all squared tree diagrams of the processes
 $e^-e^+\to W^-u\bar d$ 
and $e^-e^+\to W^+ \mu^-\bar \nu_\mu$, 
calculated in dimensional regularization without self-energy resummations, 
but expanded near threshold:
\begin{equation}
 \label{eq:four-e}
\parbox{121mm}
{\includegraphics[width=121mm]{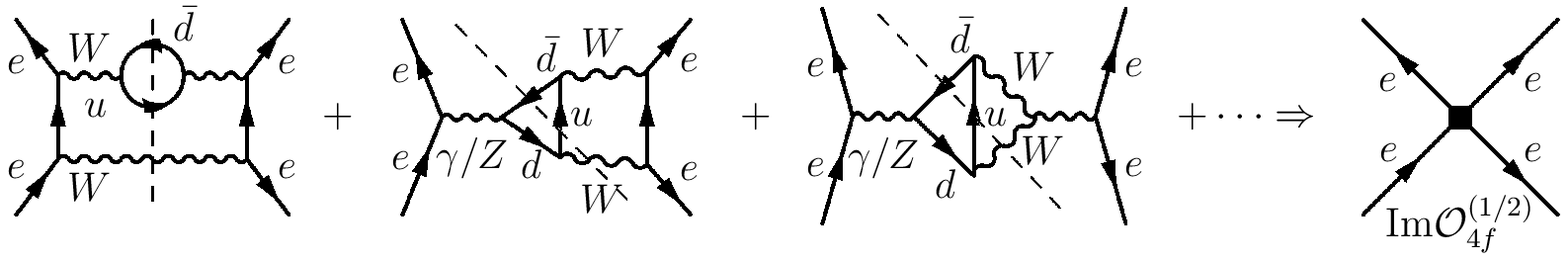}}
\end{equation}
Since these corrections to the amplitude are of order $\alpha^3$, and
counting $\alpha\sim \delta$, they are suppressed by $\delta^{1/2}$
compared to ${\cal A}^{(0)}\sim \alpha^2\delta^{1/2}$ and are denoted
as "$\sqrt \mathrm{N}$LO" corrections.

The second class of contributions arises from  \emph{production-operator and
propagator corrections}. 
Performing the tree-level
 matching~\eqref{eq:tree-match} up to order
 $\sim v$ and $v^2$ leads to higher order production operators
${\cal O}_p^{(1/2)}$ and ${\cal O}_p^{(1)}$.
The operators ${\cal O}_p^{(1/2)}$
like 
$\left(\bar{e}_L \gamma^{j} e_L \right) 
\left(\Omega_-^i (-i) D^j \Omega_+^{i}\right)$ 
are given in~\cite{Beneke:2004xd}.
At NLO one needs diagrams with two insertions of an
 ${\cal O}_p^{(1/2)}$ operator, one insertion of 
an ${\cal O}_p^{(1)}$ operator  and insertions of
kinetic corrections from~\eqref{LNR}:
\begin{equation*}
  i {\cal A}^{(1)}_{\text{born}}=
\parbox{90mm}{  \includegraphics[height=18.5mm]{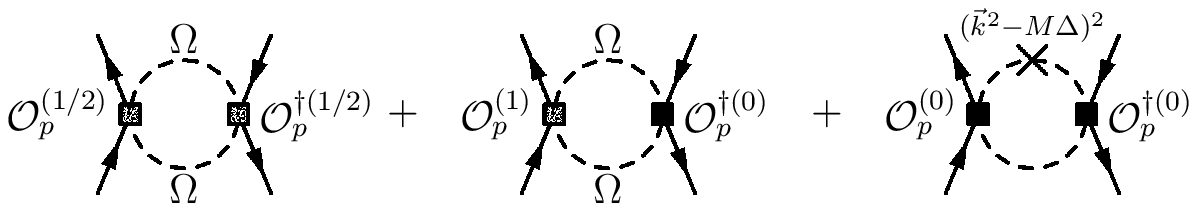}}
\end{equation*}
Equivalently one can directly expand the spin averaged
squared matrix elements~\cite{Beneke:2007zg}.

\begin{wrapfigure}{r}{7.5cm} 
\includegraphics[width=\linewidth]{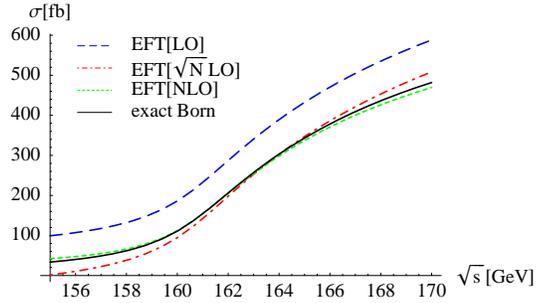}
\vspace*{-0.6cm}
\caption{Convergence of EFT approximations to the born cross section
from  Whizard}
\label{fig:eft}
\end{wrapfigure}
As seen in Figure~\ref{fig:eft}, the  
 EFT approximations converge to the full Born
result  but
it turns out that
a partial inclusion of N$^{3/2}$LO corrections 
is required to get an agreement of 
 $\sim 0.1 \%$ at 170 GeV and $\sim 10 \%$
at 155 GeV~\cite{Beneke:2007zg}. 
For higher-order initial state radiation~(ISR) improvement 
by a  convolution with radiator functions, one needs 
 $\sigma_{\text{Born}}$ at energies
 far below threshold, where the
EFT is not valid. 
For the numerical results  in Section~\ref{sec:results} 
the ISR-improved Born cross section from Whizard~\cite{Kilian:2001qz}
was used, but one could also match
the EFT to the full theory below, say, $\sqrt s =155$ GeV.

\section{Radiative corrections}
\label{sec:nlo}

The radiative corrections needed up to NLO are given by higher order
calculations of short distance coefficients and by loop calculations in the
EFT.  Counting the QCD coupling constant as $\alpha_s^2\sim \alpha_{ew}\sim
\delta$, the corrections to $\Gamma_W$ up to order
$\alpha_{ew}\alpha_s$ ($\sqrt{\text{N}}$LO), $\alpha_{ew}^2$ and
$\alpha_{ew}\alpha_s^2$ (NLO) have to be included.
  The flavour-specific NLO \emph{decay
  corrections} are correctly taken into account by
 multiplying the imaginary part of the LO
forward-scattering amplitude with the one-loop corrected branching ratios.
For the NLO \emph{renormalization of the production operator}~\eqref{LPlead}
 one has to calculate the one-loop corrections to the
on-shell scattering $e^-e^+\to W^+W^-$ at leading order in the
non-relativistic expansion:
\begin{center}
   \includegraphics[height=16mm]{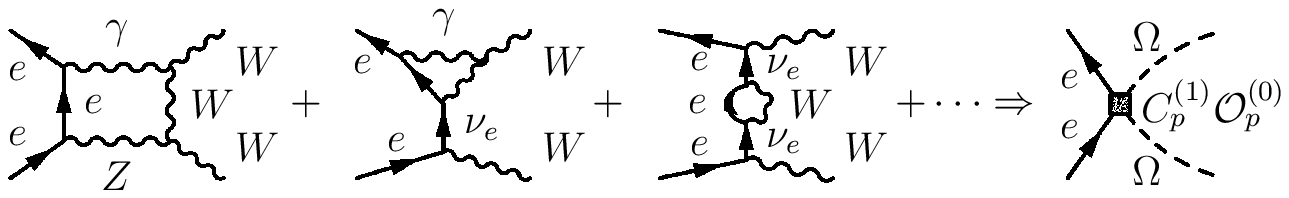}
\end{center}
 Due to the threshold kinematics, many of the 180 one-loop diagrams do
not contribute, consistent with the vanishing of
 the tree-level $s$-channel diagrams 
 at leading order in $v$.
In terms of a finite coefficient $c_{p}^{(1,{\rm fin})} $ 
 given in~\cite{Beneke:2007zg}, 
the matching coefficient reads
\begin{equation}
\label{eq:hard-matching}
C_{p}^{(1)} = \frac{\alpha}{2\pi} \left[ 
   \left(-\frac{1}{\varepsilon^2} - \frac{3}{2\varepsilon}\right)
   \left(-\frac{4 M_W^2}{\mu^2}\right)^{-\varepsilon} 
   + c_{p}^{(1,{\rm fin})} \right] \, .
\end{equation}

The \emph{first and second Coulomb correction}
arise  from the exchange of potential photons.
Their magnitude can be estimated counting
the loop-integral measure in the potential region as $d^4q\sim \delta^{5/2}$, 
the $\Omega$ propagator and the 
potential photon propagator $i/|\vec q|^2$ as $\delta^{-1}$. One finds
that single Coulomb exchange is a $\sqrt{\text{N}}$LO correction compared
to the LO amplitude:
\begin{equation} 
\parbox{22mm}{
 \includegraphics[height=15mm]{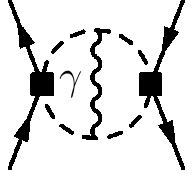}
}
\sim \alpha^3
 \int d^4 k\; d^4 q\;
  \frac{1}{|\vec q|^2}\;  \delta^{-4} 
  \;\sim\; \alpha^3 \;\sim {\cal A}^{(0)}\sqrt\delta 
\end{equation}
At threshold the one-photon exchange
is of the order of 5\% of the LO amplitude
 while two-photon exchange is only a few-permille 
correction~\cite{Fadin:1993kg} and no resummation
is necessary.

\emph{Soft photon corrections}
 correspond to two-loop
diagrams in the EFT
containing a photon with momentum $(q_0,|\vec{q}\,|) \sim (\delta,\delta)$.
They give rise to $\mathcal{O}(\alpha)$ corrections as can be seen
from a power-counting argument similar
to the one for Coulomb-exchange but 
counting the soft-photon propagator
$-i/q^2$ as 
$\delta^{-2}$ and the soft loop-integral as $\delta^4$. 
In agreement with gauge invariance arguments and
earlier  calculations~\cite{Fadin:1993dz},
the sum of all diagrams where a soft photon couples to 
an $\Omega$ line vanishes.
The only remaining diagrams give
\begin{equation}
\label{eq:soft}
\parbox{37mm}{
 \includegraphics[height=15mm]{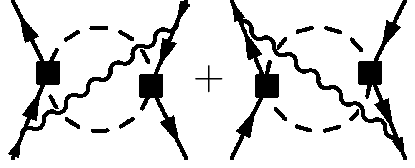}
}
\quad=\frac{4\pi^2\alpha^2}{s_w^4 M_W^2} \frac{\alpha}{ \pi}
\int \frac{d^d r}{(2 \pi)^d} \frac{1}{\eta_{-} \eta_{+}} \left[ 
  \left(\frac{1}{\varepsilon^2} +\frac{5}{12} \pi^2 \right)
   \left(-\frac{2\eta_-}{\mu}\right)^{-2\varepsilon} \right]
\end{equation}
with $\eta_{-}=r_0-\frac{|\vec{r}|^2}{2 M_W}+i\frac{\Gamma^{(0)}}{2}$ and
 $\eta_{+}=E-r_0-\frac{|\vec{r}|^2}{2
 M_W}+i \frac{\Gamma^{(0)}}{2}$. 
The $\epsilon^{-2}$ poles cancel between~\eqref{eq:soft}
and diagrams with an insertion
of the NLO production operator~\eqref{eq:hard-matching}
while the remaining $\epsilon^{-1}$ poles
 proportional to  $\left( 2 \log\left(\eta_-/M_W\right) +3/2\right)$ 
are discussed below.
\section{Results and estimate of remaining uncertainties}
\label{sec:results}

The radiative corrections in Section~\ref{sec:nlo} 
were calculated for
$m_e=0$ so the result is not infrared safe. It
should be convoluted with electron distribution
functions in the $\overline{\text{MS}}$ scheme after minimal subtraction of the
 IR poles.
However, the available distribution functions  assume 
 $m_e$ as
IR regulator. Our result can be converted to this scheme by adding
contributions from 
the  hard-collinear  
region where $k^{\mu}\sim M_W,\,k^2\sim m_e^2$, and the
soft-collinear region where $k^{\mu}\sim \Gamma_W,\,k^2\sim m_e^2 \frac{\Gamma_W}{M_W}$. 
These cancel
the $\epsilon$-poles but introduce  large logs of $M_W/m_e$:
\begin{multline}
\sigma^{(1)}(s)=\frac{\alpha^3}{27 s_w^4 s}\,\text{Im}
\Bigg\{(-1) \,\sqrt{-\tfrac{E+i \Gamma_W}{M_W}}\,
\bigg(  4\ln\bigg(-\frac{4(E+i\Gamma_W)}{M_W}\bigg)
 \ln \left(\frac{2 M_W}{m_e}\right)\\
- 5\ln \left(\frac{2 M_W}{m_e}\right)
+ \mbox{Re}\,\Big[c_{p}^{(1,\rm fin)}\Big]
+\frac{\pi^2}{4}+3\bigg)\Bigg\}
+\Delta \sigma^{(1)}_{\mbox{\tiny Coulomb}}
+\Delta \sigma^{(1)}_{\mbox{\tiny decay}}\,.
\label{eq:totcollcross}
\end{multline}

At this stage, one can compare to the results of~\cite{Denner:2005es}
for the strict $\mathcal{O}(\alpha_{ew})$ corrections without higher
order ISR improvement, $\sigma_{4\text{f}}(161\text{GeV}) = 105.71(7)$
fb and $\sigma_{4\text{f}}(170\text{GeV}) = 377.1(2)$.
From~\eqref{eq:totcollcross} one obtains
$\sigma_{\text{EFT}}(161\text{GeV}) =104.97(6)$ fb and
$\sigma_{\text{EFT}}(170\text{GeV}) = 373.74(2)$ so
 the difference between the EFT and~\cite{Denner:2005es}
is only about  $0.7\%- 1\%$.

\begin{wrapfigure}{R}{6cm} 
\includegraphics[width=1.0\linewidth]{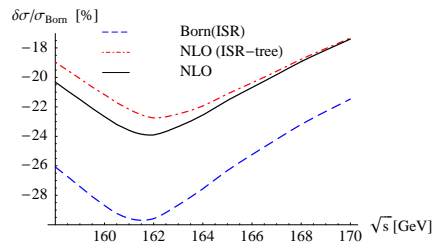}
\vspace*{-0.6cm}
\caption{Size of the relative NLO corrections for
 different treatments of ISR}
\label{fig:relcorr}
\end{wrapfigure}

The large logs in~\eqref{eq:totcollcross}
can be resummed by convoluting the NLO cross section
with the structure functions
used e.g. in~\cite{Denner:1999kn}, 
after appropriate subtractions
to avoid double-counting.
The solid line in Figure~\ref{fig:relcorr} 
shows the resulting corrections 
relative to $\sigma_{\text{Born}}$. Compared to the
large correction from ISR improvement of
$\sigma_{\text{Born}}$ alone (blue/dashed),
the size
of the genuine radiative correction is
about $+8\%$.

The largest remaining uncertainty 
is due to the 
treatment of ISR that is accurate only at 
leading-log level. 
It is formally equivalent to
 improve only $\sigma_{\text{Born}}$ by higher order ISR~\cite{Denner:1999kn,Denner:2005es},  but not 
the radiative corrections. 
The results of this approach are
 shown in the red (dash-dotted) line in Figure~\ref{fig:relcorr} and
differ by almost $2\%$ at threshold from the treatment
 discussed above. 
This translates to an 
 uncertainty of
 $\delta M_W\sim 31\, \mbox{MeV}$~\cite{Beneke:2007zg}.
The remaining theory uncertainty comes from the uncalculated N$^{3/2}$LO
 corrections in the EFT. 
The $O(\alpha)$
corrections to the the four-electron
 operators~\eqref{eq:four-e} 
lead to an estimated  uncertainty
of $ \delta M_W\sim 8$ MeV~\cite{Beneke:2007zg}.
These corrections 
are included in~\cite{Denner:2005es}.
The effect of diagrams with single-Coulomb
exchange together with a soft photon or a hard correction to the production
vertex is  estimated as
$\delta M_W\sim 5$ MeV.
Therefore it should be possible
to reach the theoretical accuracy  required  for
the $M_W$ measurement
since the largest remaining uncertainties can be eliminated by
an improved treatment of
ISR and with input 
of the full four fermion calculation.

\section*{Acknowledgments}
I thank M. Beneke, P. Falgari, A. Signer and G. Zanderighi for
the collaboration on~\cite{Beneke:2007zg}
and for comments on the manuscript.
 I acknowledge support by the DFG SFB/TR~9.

\providecommand{\href}[2]{#2}
\begin{footnotesize}


\end{footnotesize}

\end{document}